\documentclass[twocolumn,showpacs,preprintnumbers,amsmath,amssymb]{revtex4}

\usepackage{graphicx}
\usepackage{textcomp}

\begin{document}
\draft
\title{$\mathcal{PT}$-symmetric laser-absorber}
  \normalsize

\author{Stefano Longhi}
\address{Dipartimento di Fisica, Politecnico di Milano, Piazza L. da Vinci 32, I-20133 Milano, Italy}


%
\bigskip
\begin{abstract}
\noindent In a recent work, Y.D. Chong et al. [Phys. Rev. Lett. {\bf
105}, 053901 (2010)] proposed the idea of a coherent perfect
absorber (CPA) as the time-reversed counterpart of a laser, in which
a purely incoming radiation pattern is completely absorbed by a
lossy medium. The optical medium that realizes CPA is obtained by
reversing the gain with absorption, and thus it generally differs
from the lasing medium. Here it is shown that a laser with an
optical medium that satisfies the parity-time $(\mathcal{PT})$
symmetry condition $\epsilon(-\mathbf{r})=\epsilon^*(\mathbf{r})$
for the dielectric constant behaves simultaneously as a laser
oscillator (i.e. it can emit outgoing coherent waves) and as a CPA
(i.e. it can fully absorb incoming coherent waves with appropriate
amplitudes and phases). Such a device can be thus referred to as a
$\mathcal{PT}$-symmetric CPA-laser. The general
amplification/absorption features of the $\mathcal{PT}$ CPA-laser
below lasing threshold driven by two fields are determined.
\end{abstract}


\maketitle

{\it Introduction.} It is well known that a gain medium embedded in
an optical cavity, i.e. a laser oscillator, emits a coherent
electromagnetic radiation that escapes from the cavity in the form
of outgoing monochromatic waves when the amplification of photons in
the medium reaches a threshold value that balances light leakage out
of the cavity \cite{Siegman}. In a recent work \cite{Chong10}, Y.D.
Chong and collaborators have introduced the concept of a coherent
perfect absorber (CPA) as the time-reversed counterpart of a laser,
in which by reversing the gain with absorption the same optical
system supports a purely incoming radiation pattern -the
time-reverse of the lasing mode- with complete absorption and zero
reflection. Generally, the laser medium and its time-reversed medium
are different media because
gain must be replaced with loss.\\
In this Report it is shown that an optical medium that satisfies the
parity-time $(\mathcal{PT})$ symmetry condition
$\epsilon(-\mathbf{r})=\epsilon^*(\mathbf{r})$ for the dielectric
constant can behave {\em simultaneously} as a laser oscillator (i.e.
it can emit outgoing coherent waves) and as a coherent perfect
absorber, fully absorbing incoming coherent waves with the right
amplitudes and phases. Owing to such a rather special property, we
refer such an optical device to as a $\mathcal{PT}$ CPA-laser.
Optical structures with a complex refractive index with even (odd)
symmetry for the real (imaginary) part of the refractive index
belong to the so-called $\mathcal{PT}$ optical structures
\cite{PT1,PT2}. Such structures have received in the past few years
an increasing attention in the optical and physical communities
\cite{PT1,PT2}, mainly because they provide an experimentally
accessible laboratory system to simulate with optics certain
features of $\mathcal{PT}$-symmetric and pseudo-Hermitian quantum
mechanics \cite{Bender,MostafazadehQM}. A $\mathcal{PT}$ CPA-laser
can be realized using a $\mathcal{PT}$-invariant distributed
feedback optical structure, which was recently proposed in the
framework of $\mathcal{PT}$ relativistic quantum mechanics \cite{Longhi10}.\\
\\
{\it The idea of the $\mathcal{PT}$-symmetric CPA-laser.} The main
idea underlying a $\mathcal{PT}$ CPA-laser can be captured by
considering monochromatic wave propagation in a dielectric structure
with a spatially-dependent complex dielectric constant $\epsilon$,
that realizes the laser oscillator, in the plane-wave and scalar
approximations. Such approximations hold, for example, for a laser
oscillator realized in a fiber or in a waveguide. By writing the
electric field in the structure as $\mathcal{E}(x,t)=E(x) \exp(-i
\omega t)+c.c.$, where $\omega$ is frequency of the field, the
spatial field envelope $E(x)$ satisfies the Helmholtz equation
\begin{equation}
\frac{d^2E}{dx^2}+\left(\frac{\omega}{c_0}\right)^2 \epsilon(x)E=0
\end{equation}
where $c_0$ is the speed of light in vacuum. In general, the
frequency $\omega$ is permitted to assume complex values; in this
case the imaginary part of $\omega$ corresponds to the growth rate
in time of the spatial mode $E(x)$. The lasing medium is assumed to
be embedded in the spatial region $|x|<L/2$, where $\epsilon(x)$ is
complex-valued. The imaginary part of $\epsilon$ describes the local
gain or loss of the medium. Outside the laser oscillator, i.e. for
$|x|>L/2$, $\epsilon(x)$ is assumed to be real-valued  and equal to
$\epsilon(x)=n_0^2$, where $n_0$ is the (modal) refractive index of
the waveguide or fiber [see Fig.1(a)]. For a laser at threshold,
Eq.(1) admits of a solution $E=E_0(x)$ describing outgoing waves at
some real frequency $\omega=\omega_0$ (the frequency of the most
unstable mode), i.e. with the behavior $E_0(x)= \exp(ikn_0x)$ for
$x>L/2$ and $E_0(x)= q \exp(-ikn_0x)$ for $x<-L/2$, where
$k=\omega/c_0$ and $q$ is a constant describing the unbalance of the
outgoing waves from the two sides of the cavity. If we take the
complex conjugate of both sides of Eq.(1), use the relation
$\epsilon(-x)=\epsilon^*(x)$ and make the spatial inversion $x
\rightarrow -x$, it readily follows that $E_1(x)=E^*_0(-x)$ is also
a solution to Eq.(1) with the same frequency $\omega=\omega_0$ and
dielectric constant $\epsilon(x)$. Such a new solution has the
asymptotic behavior $E_1(x)= q^* \exp(-ikn_0x)$ for $x>L/2$ and
$E_1(x)= \exp(ikn_0x)$ for $x<-L/2$, i.e. it corresponds to two
incoming waves that are fully absorbed in the medium, without being
reflected. Thus a laser at threshold for which the dielectric
constant satisfies the $\mathcal{PT}$ symmetry condition can {\it
simultaneously} generate outgoing monochromatic waves and perfectly
absorb incoming waves with appropriate amplitude and phase. Further
properties of the $\mathcal{PT}$ CPA-laser can be gained by a more
detailed
analysis of the scattering properties of the optical structure.\\
\\
{\it Scattering analysis and $\mathcal{PT}$ CPA-laser.} The most
general solution to Eq.(1) has the asymptotic behavior $E(x)=a
\exp(ikn_0x)+b \exp(-ikn_0x)$ for $x<-L/2$, and $E(x)=c
\exp(ikn_0x)+d \exp(-ikn_0x)$ for $x>L/2$. The amplitudes of
forward- and backward-propagating waves out of the cavity are
related by the algebraic relation
\begin{equation}
\left(
\begin{array}{c}
c \\ d
\end{array}
\right) =\mathcal{M}(\omega) \left(
\begin{array}{c}
a \\ b
\end{array}
\right)
\end{equation}
where $\mathcal{M}(\omega)$, with ${\rm det} \mathcal{M}=1$, is the
$2 \times 2$ transfer matrix of the optical structure from $x=-L/2$
to $x=L/2$ [see Fig.1(a)]. The spectral transmission and reflection
coefficients for left ($l$) and right ($r$) incidence, i.e. the
coefficients of the scattering matrix, can be expressed in terms of
the transfer matrix elements as
\begin{equation}
t^{(l)}=t^{(r)}\equiv t =\frac{1}{\mathcal{M}_{22}}, \;
r^{(l)}=-\frac{\mathcal{M}_{21}}{\mathcal{M}_{22}} ,  \;
r^{(r)}=\frac{\mathcal{M}_{12}}{\mathcal{M}_{22}}.
\end{equation}
For a laser oscillator without injected signal, the boundary
conditions $a=d=0$ apply, which imply [from Eq.(2)]
$\mathcal{M}_{22}(\omega)=0$. If all the zeros of
$\mathcal{M}_{22}(\omega)$ (i.e. the poles of the transmission $t$)
lie in the lower half of the imaginary plane, i.e. $\mathrm{
Im}(\omega)<0$, the laser is below threshold. As the gain is
increased, the poles of $t$ move toward the real axis, and laser
threshold is reached when the most unstable pole, say
$\omega=\omega_0$, becomes real. On the other hand, for a perfect
absorber the boundary conditions $b=c=0$, i.e. the absence of
reflected waves, hold. From Eq.(2) this implies
$\mathcal{M}_{11}(\omega)=0$, and the amplitudes of incident waves
must satisfy the condition $d=\mathcal{M}_{21}(\omega) a$ . For a
laser at threshold, for which $\mathcal{M}_{22}(\omega_0)=0$ at some
real frequency $\omega_0$, the condition
$\mathcal{M}_{11}(\omega)=0$ is generally never satisfied at any
frequency $\omega$, i.e. a laser {\em does not generally behave}
like a perfect absorber. For example, let us consider a standard
laser model comprising two equal lossless end mirrors, at $x= \pm
L/2$, filled by a gain medium with gain coefficient $g$ [Fig.1(b)].
The transfer matrix $\mathcal{M}$ of this simple laser oscillator is
given by the product of the transfer matrices of the mirrors and
gain medium, i.e. $\mathcal{M}=\mathcal{M}_M \mathcal{M}_g
\mathcal{M}_M$, where \cite{Siegman,Haus}
\begin{eqnarray}
\mathcal{M}_M & = & \frac{1}{t_M} \left(
\begin{array}{cc}
t_M^2-r_M^2 & r_M \\
-r_M & 1
\end{array}
\right) \nonumber \\
\mathcal{M}_g & = & \left(
\begin{array}{cc}
\exp(gL+i \delta) & 0 \\
0 & \exp(-gL-i \delta)
\end{array}
\right),
\end{eqnarray}
 $\delta=(\omega/c_0)n_g(\omega)L$,
$n_g(\omega)$ is the real part of the refractive index in the gain
medium, and $r_M$, $t_M$ are the field reflection and transmission
coefficients of the mirrors. For a lossless passive mirror, one has
$|t_M|^2+|r_M|^2$=1 and the phases of $r_M$ and $t_M$ differ by $\pm
\pi/2$ (see, for instance, \cite{Haus}). For the sake of
definiteness, we assume $r_M$ to be real-valued, so that $t_M$ is
purely imaginary. After straightforward calculations, the condition
$\mathcal{M}_{22}(\omega)=0$ for a real frequency $\omega$ yields
$\exp(-2gL-2i\delta)=r_M^2$, i.e. $\exp(-2gL)=r_M^2$ and $\delta=l
\pi$ ($l$ is an integer number). The former equation corresponds to
the usual threshold condition for laser oscillation, stating the
balance of coupling losses with gain in the medium; the latter
equation states that the lasing frequency should correspond to one
among the longitudinal resonance frequencies of the resonator
\cite{Siegman}. On the other hand, the condition for a perfect
absorber, $\mathcal{M}_{11}=0$, reads explicitly $r_M^2 \exp(-2gL-2i
\delta)=1$, which can never be satisfied for any real frequency
$\omega$. Hence an ordinary laser oscillator does not behave as a perfect absorber.\\
 Let us consider now a $\mathcal{PT}$-symmetric laser oscillator; a possible
 realization of such an oscillator will be discussed below. In this case, taking the complex conjugate of Eq.(1), using the
relation $\epsilon(-x)=\epsilon^*(x)$ and after setting $x
\rightarrow -x$, from Eq.(2) it ca be readily shown that
$\mathcal{M}^{-1}(\omega)=\mathcal{M}^*(\omega^*)$, i.e.
\begin{eqnarray}
\mathcal{M}_{22}(\omega) & = & \mathcal{M}_{11}^*(\omega^*) \\
\mathcal{M}_{12}(\omega)& = & -\mathcal{M}_{12}^*(\omega^*), \;
\mathcal{M}_{21}(\omega)=-\mathcal{M}_{21}^*(\omega^*).
\end{eqnarray}
Therefore, if the medium is at threshold for lasing, there exists a
real frequency $\omega_0$ such that $\mathcal{M}_{22}(\omega_0)=0$.
Form Eq.(5), it follows that at the same frequency one has
$\mathcal{M}_{11}(\omega_0)=0$, i.e. the lasing medium also behaves
as a CPA, according to the simple argument presented in the previous section.\\
\begin{figure}
\includegraphics[scale=0.7]{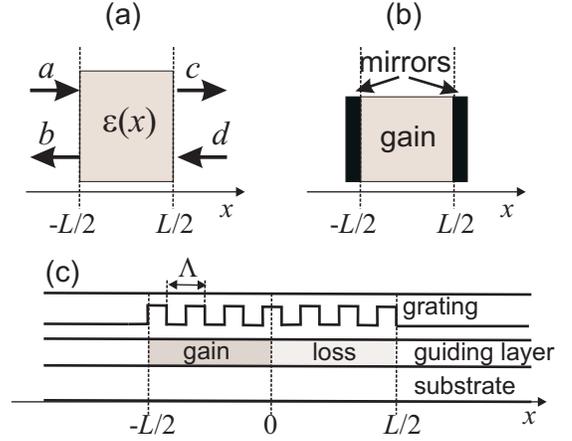}
\caption{(a) Schematic of wave scattering in a one-dimensional
optical structure with complex dielectric constant $\epsilon(x)$.
(b) Schematic of a laser oscillator comprising two equal lossless
mirrors filled by a gain medium. (c) Distributed-feedback structure,
consisting of a uniform index grating with two homogeneous and
symmetric gain and loss regions, that realizes a $\mathcal{PT}$
CPA-laser.}
\end{figure}
\\
{\em Amplification and absorption properties of the $\mathcal{PT}$
CPA-laser below threshold}. It is interesting to investigate the
amplification and absorption properties of the
$\mathcal{PT}$-symmetric laser just below threshold for oscillation
\cite{note1} when two signals, of amplitudes $a$ and $d$ [see
Fig.1(a)], are injected from the two sides of the cavity, i.e. when
the laser is operated as a cavity amplifier with two input ports
(see, for instance, \cite{Mandel89}). Near threshold and for
frequencies $\omega$ of the injected signals close to the laser
frequency $\omega_0$, the most general form of the coefficients of
the transfer matrix $\mathcal{M}(\omega)$, satisfying the constrain
Eqs.(5) and (6) imposed by the $\mathcal{PT}$ invariance, read
\begin{eqnarray}
\mathcal{M}_{11} & = & \kappa (\omega-\omega_0-i \epsilon) \\
\mathcal{M}_{22} & = & \kappa^* (\omega-\omega_0+i \epsilon) \\
\mathcal{M}_{12} & = & i \alpha +i \beta (\omega-\omega_0) \\
\mathcal{M}_{21} & = &
\frac{\mathcal{M}_{11}\mathcal{M}_{22}-1}{\mathcal{M}_{12}}
 \simeq \frac{i}{\alpha} -i \frac{\beta}{\alpha^2} (\omega-\omega_0) +
 o(\epsilon) \;\;\;\;\;\;\;\;\;
\end{eqnarray}
where $\epsilon>0$ is a small parameter that measures the distance
from laser threshold, $\alpha$ and $\beta$ are real-valued
parameters, and $\kappa$ a complex parameter. In writing Eqs.(7-10),
we assumed that $\mathcal{M}_{22}$ has a simple zero at
$\omega=\omega_0-i\epsilon$ (i.e. $t=1/\mathcal{M}_{22}$ has a
simple pole), and $\omega-\omega_0 \sim \epsilon$. For a single
injected signal at frequency $\omega$, for example for $d=0$ (left
incident signal), the $\mathcal{PT}$ laser behaves like an ordinary
cavity amplifier \cite{note1}, and the spectral intensity
transmission
$T=|t(\omega)|^2=1/|\kappa^2|[(\omega-\omega_0)^2+\epsilon^2]$ has a
typical Lorentzian profile. For the following discussion, it is
worth introducing an overall reflection/transmission coefficient
$\Theta$, defined as the ratio of the total intensity of outgoing
(reflected/transmitted) waves over the total intensity of the
incoming (injected) waves \cite{Chong10}, i.e.
\begin{equation}
\Theta=\frac{|b|^2+|c|^2}{|a|^2+|d|^2}.
\end{equation}
Note that the vanishing of $\Theta$ is the signature of perfect
absorption, whereas $\Theta>1$ indicates that an overall
amplification has been realized in the medium \cite{note}. For a
single-port injection ($d=0$), one has
$\Theta=(|b|^2+|c|^2)/|a|^2=T+R^{(l)}$, where $R^{(l)}=|r^{(l)}|^2$
is the spectral intensity reflection for left incidence. A typical
behavior of $\Theta(\omega)$ in this case is shown in Fig.2(a),
dotted curve. Note that, because $\Theta>T$, near the resonance,
$\omega=\omega_0$, $\Theta(\omega)$ takes large values, diverging as
the lasing threshold is approached (i.e. $\epsilon \rightarrow 0$).
It is now interesting to investigate how the second signal of
amplitude $d$, injected into the other side of the cavity, can fully
change the behavior of $\Theta(\omega)$, and in particular how
perfect absorption can be attained in a narrow spectral range at
around $\omega=\omega_0$ for appropriate amplitude and phase of $d$.
To this aim, let us first consider a coherent signal and set
$d/a=\sigma \exp(i \phi)$; in this case one obtains
\begin{equation}
\Theta(\omega)=\frac{|1+\sigma \mathcal{M}_{12}\exp(i
\phi)|^2+|\sigma \exp(i
\phi)-\mathcal{M}_{21}|^2}{(1+\sigma^2)|\mathcal{M}_{22}|^2}.
\end{equation}
To achieve CPA, according to the previous scattering analysis let us
assume a second coherent signal with amplitude and phase defined by
$d/a=\sigma \exp(i \phi)=\mathcal{M}_{21}(\omega_0)$. In this case,
from Eq.(12) one obtains $\Theta(\omega_0) \sim \epsilon^2$, i.e.
perfect absorption is attained at $\omega=\omega_0$ as $\epsilon
\rightarrow 0$. A typical behavior of $\Theta(\omega)$ in this case,
depicted in Fig.2(a) (solid curve), shows a marked dip near
$\omega=\omega_0$. Thus the second injected signal, of appropriate
amplitude and phase, makes the amplifier a (near) perfect absorber
in a narrow spectral range around $\omega_0$. It should be noted
that such perfect absorption requires that the two injected fields
be coherent. For example, let us assume that the two fields have a
definite amplitude ratio $\sigma=|d/a|$, however their relative
phase $\phi$ fluctuates. Averaging the expression of
$\Theta(\omega)$ over the fluctuating phase $\phi$ yields
\begin{equation}
\Theta(\omega)=\frac{1+\sigma^2+\sigma^2
|\mathcal{M}_{12}|^2+|\mathcal{M}_{21}|^2}{(1+\sigma^2)|\mathcal{M}_{22}|^2}.
\end{equation}
As in the CPA case, let us assume
$\sigma=|\mathcal{M}_{21}(\omega_0)|$. A typical behavior of
$\Theta(\omega)$ for incoherent excitation is shown by the dashed
curve in Fig.2(a). Note that, similarly to the single injected
signal, in this case $\Theta$ diverges (rather than vanishing) at
$\omega=\omega_0$ as the lasing threshold is attained. This
indicates that, as in \cite{Chong10}, CPA requires both an amplitude
and phase control of the two incoming waves. Therefore, depending on
the amplitude and phase of the two injected signals, the
$\mathcal{PT}$ CPA-laser below threshold can either amplify or
absorb the input fields. At exact threshold, the $\mathcal{PT}$
CPA-laser behaves like and ordinary laser, i.e. outgoing waves will
be spontaneously emitted starting from spontaneous emission noise in
the medium. However, as compared to an ordinary laser, the
$\mathcal{PT}$ CPA-laser has the additional property to fully absorb
incoming radiation from the two ports with appropriate amplitude and
phase relationship.
\\
\begin{figure}
\includegraphics[scale=0.48]{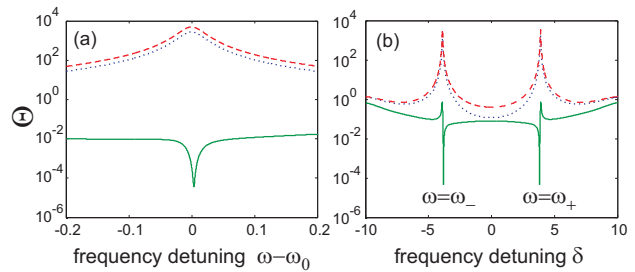}
\caption{(color online) (a) Behavior of the overall
reflection/transmission coefficient $\Theta$ (on a logarithmic
scale) versus frequency detuning $\omega-\omega_0$ for the transfer
matrix defined by Eqs.(7-10) and for parameter values $\alpha=3$,
$\beta=0.3$, $\epsilon=0.02$, and $\kappa=1$. Solid curve: two-port
coherent input excitation with $d/a=\mathcal{M}_{21}(\omega_0)$;
dashed curve: two-port incoherent input excitation with
$d/a=|\mathcal{M}_{21}(\omega_0)| \exp(i \phi)$, averaged over the
random phase $\phi$; dotted curve: single-port excitation ($d=0$).
(b) Same as (a), but for the distributed-feedback structure of
Fig.1(c) for parameter values $q_0L=1$ and $gL=4.43$. The normalized
frequency detuning in the horizontal axis is defined as
$\delta=(\omega-\omega_B)n_0L/c_0$, where $\omega_B$ is the Bragg
reference frequency. }
\end{figure}
\\
{\it Distributed-feedback $\mathcal{PT}$ CPA-laser.} How one can
realize a $\mathcal{PT}$-symmetric CPA-laser? At first sight one
could naively think that, because the overall (spatially-averaged)
gain coefficient in a $\mathcal{PT}$-invariant structure always
vanishes, a $\mathcal{PT}$ laser would never reach threshold for
oscillation. However, this is not the case and threshold for laser
oscillation can be generally reached at the so-called $\mathcal{PT}$
symmetry-breaking point \cite{Longhi10}. At such a point, a couple
of resonance/antiresonance crosses the real frequency axis
\cite{Longhi10}, defining the lasing and CPA modes of the CPA-laser
system. The photon in the former (lasing) mode lives more in the
gain part of the structure, while the photon in the latter (CPA)
mode lives more in the lossy part. A possible way to realize a
$\mathcal{PT}$ CPA-laser with a reasonable low threshold is to
consider a distributed-feedback structure, composed by a uniform
index grating with refractive index $n(x)=n_0+\Delta n \cos(2 \pi
x/\Lambda)$, with two homogeneous gain and lossy regions at
$-L/2<x<0$ and $0<x<L/2$, respectively, with a gain/loss coefficient
per unit length equal to $g$ [Fig.1(c)]. Such a structure has been
recently introduced in Ref.\cite{Longhi10} as an optical system to
test the onset of spectral singularities and $\mathcal{PT}$ symmetry
breaking in an optical analogue of the non-Hermitian relativistic
Dirac equation.  As $g$ is increased from zero, threshold for laser
oscillation is attained at two frequencies $\omega_{\pm}=\omega_B
\pm \Delta \omega$, symmetrically placed at around the Bragg
frequency $\omega_B=\pi c_0/\Lambda$ (see, for instance, Fig.2(c) of
Ref.\cite{Longhi10}). The transfer matrix of the
distributed-feedback structure is given by \cite{Longhi10}
$\mathcal{M}=\mathcal{M}^{+}\mathcal{M}^-$, where
\begin{eqnarray}
\mathcal{M}^{\pm}_{11} & = &
\cosh(\lambda_{\pm}L)-i\frac{\rho_{\pm}}{\lambda_{\pm}}
\sinh(\lambda_{\pm}L) \\
\mathcal{M}^{\pm}_{12} & = &
 -i\frac{q_0}{\lambda_{\pm}}
\sinh(\lambda_{\pm}L) \; , \; \; \mathcal{M}^{\pm}_{21}  =
i\frac{q_0}{\lambda_{\pm}}
\sinh(\lambda_{\pm}L) \;\;\;\;\;\;\;\;\;\;\; \\
\mathcal{M}^{\pm}_{22} & = &
\cosh(\lambda_{\pm}L)+i\frac{\rho_{\pm}}{\lambda_{\pm}}
\sinh(\lambda_{\pm}L)
\end{eqnarray}
In Eqs.(14-16), we have set $q_0=\omega_B \Delta n/(2c_0)$,
$\rho_{\pm}=n_0(\omega-\omega_0)/c_0 \pm i g$, and
$\lambda_{\pm}=\sqrt{q_0^2-\rho_{\pm}^2}$.
 The threshold gain
value $g_{th}L$ for laser oscillation can be found numerically by
searching for the zeros of $\mathcal{M}_{22}$; in particular,
$g_{th}L$ turns out to be a function of $q_0L$ solely. For example,
for $q_0L=1$, one has $g_{th}L \simeq 4.46$. When the structure is
kept below lasing threshold and is illuminated with a single field,
or with two fields -either coherent or incoherent- with amplitude
ratio
 $|d/a|=|\mathcal{M}_{21}(\omega_{\pm})|$, the numerically-computed behavior of $\Theta$
versus the normalized frequency detuning
$\delta=(\omega-\omega_B)n_0L/c_0$ is shown in Fig.2(b) for
$gL=4.43$. Note that, for coherent signal injection CPA is observed
near the lasing frequencies $\omega_{\pm}$, according to the general
analysis developed in the previous section. It is worth observing
the the dips of the CPA near $\omega_{\pm}$ are much narrower than
the peaks of the amplifier when the $\mathcal{PT}$ system is
operated with a single input signal or with two incoherent signals
[see Fig.2(b)].\\
\\
{\em Conclusions.} In Ref.\cite{Chong10}, the idea that backwards
lasing yields a coherent perfect absorber has been proposed. The
optical media that realize the laser oscillator and the CPA are
generally different each other. Here we have introduced the idea of
a $\mathcal{PT}$-symmetric CPA-laser, and shown that an optical
medium that satisfies the $\mathcal{PT}$ symmetry condition
$\epsilon(-\mathbf{r})=\epsilon^*(\mathbf{r})$ can behave
simultaneously as a laser oscillator, emitting outgoing coherent
waves, and as a coherent perfect absorber, fully absorbing incoming
coherent waves with appropriate amplitudes and phases. The
amplification and absorption properties of the $\mathcal{PT}$
CPA-laser below lasing threshold have
been also discussed.\\
\\
Work supported by the italian MIUR (Grant No. PRIN-2008-YCAAK,
"Analogie ottico-quantistiche in strutture fotoniche a guida
d'onda"). Fruitful discussions and physical insights by Prof. A.
Douglas Stone are gratefully acknowledged.


\begin{thebibliography}{31}
\bibitem{note}
For a lossless passive medium [i.e. $\epsilon(x)$ real-valued] one
always has $\Theta(\omega)=1$ for power conservation.
\end{thebibliography}

\end{document}